\documentclass[twocolumn,prl,superscriptaddress,showpacs]{revtex4}
\usepackage{epsfig}
\usepackage{times}

\begin{document}

\title{Probing the pairing symmetry and pair charge stiffness of doped $t-J$ ladders} 

\author{A.~E.~Feiguin}
\affiliation{Microsoft Research, Project Q, University of California, Santa Barbara, CA 93106}
%\email{afeiguin@microsoft.com}
\author{S.~R.~White}
\affiliation{Department of Physics and Astronomy, University of California, Irvine, CA 92697}
%\email{srwhite@uci.edu}
\author{D.~J.~Scalapino}
\affiliation{Department of Physics, University of California, Santa Barbara, CA 93106-9530}
%\email{djs@vulcan2.physics.ucsb.edu}
\author{I.~Affleck}
\affiliation{Department of Physics and Astronomy, University of British Columbia, Vancouver, British Columbia, Canada V6T 1Z1}
%\email{iaffleck@phas.ubc.ca}

\date{\today}

\begin{abstract}
We perform the numerical equivalent of a phase sensitive experiment on doped $t-J$ ladders. 
We apply proximity
effect fields with different complex phases at both ends of an open system
 and we study the transport of Cooper pairs.
 Measuring the response of the system and the induced Josephson current, 
Density Matrix Renormalization Group calculations show how, depending on the 
doping fraction, 
the rung-leg parity of the pair field changes from minus to plus as the density of holes is increased. We also study the pair charge stiffness, and we observe a supression of the superconductivity in the region where static stripes appear.
We compare our results with predictions from bosonization and renormalization group analysis.
\end{abstract}


\maketitle

%INTRODUCTION
%METHOD
%RESULTS

\begin{centering}
\begin{figure}
\epsfig {file=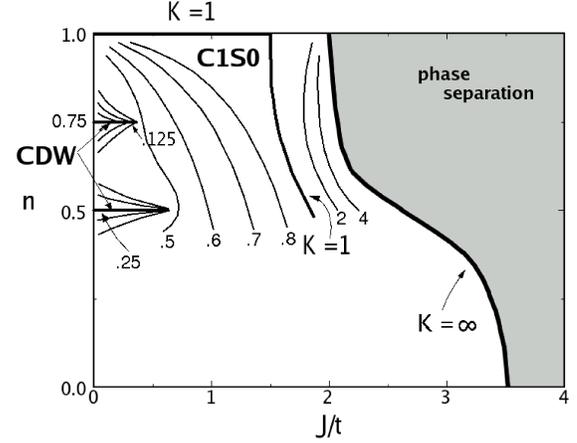,width=80mm}
\caption{Schematic phase diagram of the $t-J$ ladder as a function of $J/t$ and site density $\langle n \rangle$, from Refs.[\onlinecite{friedel,phase diag}]. 
}
\label{fig1}
\end{figure}
\end{centering}

The 2-leg $t-J$ ladder provides an example of how an apparently simple model of a strongly correlated many electron system can exhibit a rich variety of phenomena.\cite{tj ladder} This is illustrated by the schematic phase diagram reproduced \cite{friedel,phase diag} in Fig.~\ref{fig1}. Here $J/t$ is the ratio of the near neighbor exchange $J$ to the near neighbor hopping $t$, and $\langle n \rangle$ is the site filling. For $\langle n \rangle = 1$, one has a 2-leg Heisenberg ladder which has a spin gap $\Delta_s \approx J/2$. When holes are added, the system enters a spin gapped `d-wave-$4k_F$CDW' Luther-Emery \cite{Luther and Emery} phase for physical values of $J/t$. In this region there are power law pairfield and $4k_F$-CDW correlations. For $\langle n \rangle \ge 0.5$, the pairfield correlations are said to be `d-wave' like because the correlation function of the singlet rung and singlet near neighbor leg pairfields is negative. For larger, unphysical values of $J/t$, the system phase separates. Density Matrix Renormalization Group  calculations\cite{friedel} also provide evidence of a commensurate CDW phase at smaller values of $J/t$ for $\langle n \rangle = 0.75$ and $0.5$. At larger doping, where $\langle n \rangle < 0.5$, there is a gapless Luttinger liquid phase and at low doping and large $J/t$ values, an electron pairing region characterized by `s-wave'-like positive rung-leg pairfield correlations.

\begin{centering}
\begin{figure}
\epsfig {file=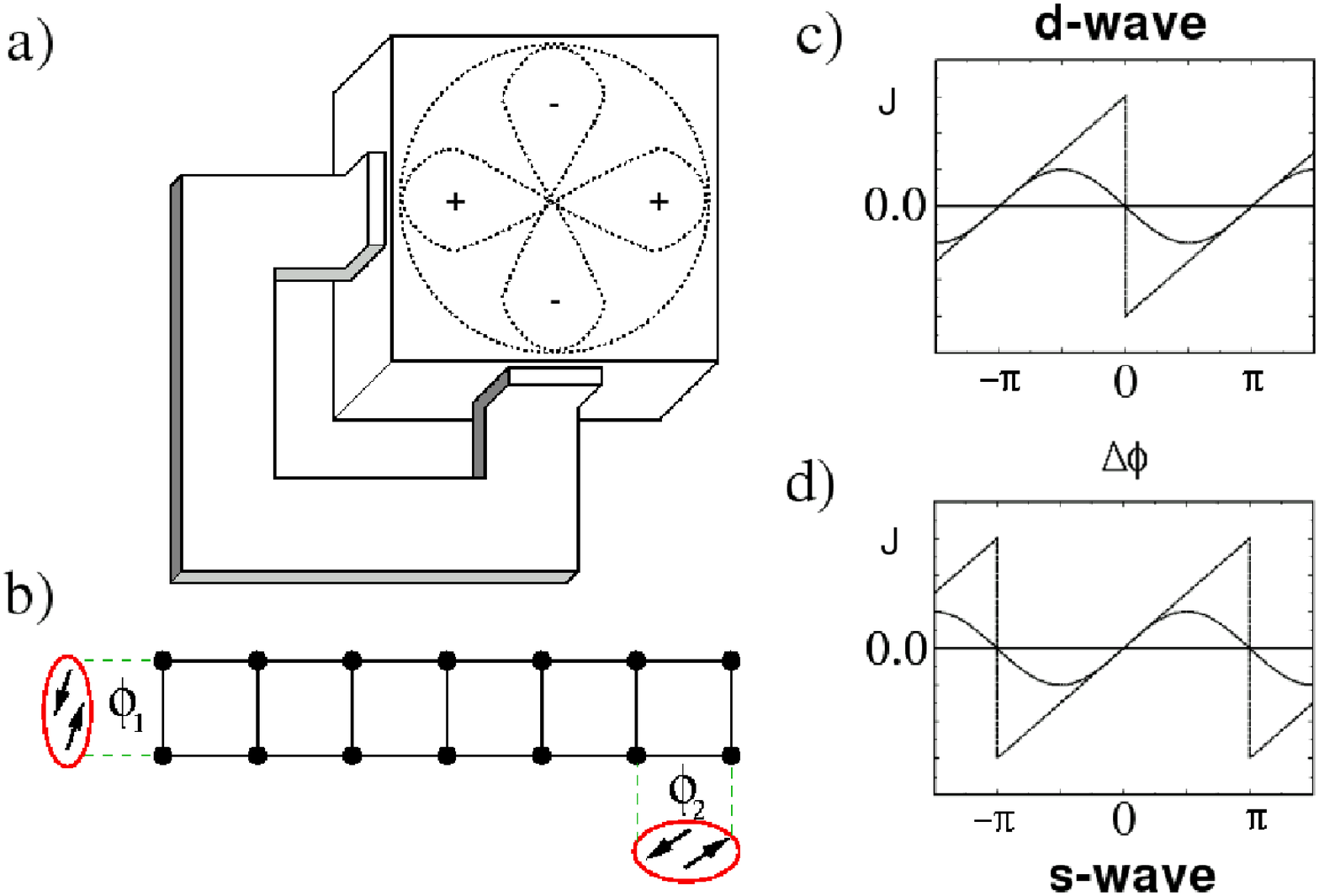,width=80mm}
\caption{a) Corner configuration used in SQUID phase sensitive experiments, from Ref.\cite{Van Harlingen}; b) Ladder geometry used in our calculations; c) and d) Typical current response expected from d-wave and s-wave like superconducting orders, respectively.
The sawtooth profile corresponds to perfect Andreev reflection, while the sine wave response describes the behavior as Andreev reflection becomes small.}
\label{fig2}
\end{figure}
\end{centering}

Here we investigate the pairfield response of the 2-leg $t-J$ ladder using a recently developed numerical technique\cite{SHS} in which pairfields are applied to the ends of the ladder, as illustrated in Fig.~\ref{fig2}b). The Hamiltonian for the ladder has the usual form
\begin{eqnarray}
H = & - & t\sum_{i,\lambda,\sigma}\left(c^\dagger_{i,\lambda\sigma} c^{\phantom{\dagger}}_{i+1,\lambda\sigma}+h.c.\right)\nonumber\\
& - & t\sum_{i,\sigma}\left(c^\dagger_{i2\sigma} c^{\phantom{\dagger}}_{i1\sigma}+h.c.\right)\nonumber\\
& + & J\sum_{i,\lambda}\left(\vec{S}_{i+1,\lambda} \cdot \vec{S}_{i,\lambda}-\frac{n_{i+1,\lambda}n_{i\lambda}}{4}\right)\nonumber\\ 
& + & J\sum_{i}\left(\vec{S}_{i1} \cdot \vec{S}_{i2}-\frac{n_{i1}n_{i2}}{4}\right) 
\end{eqnarray} 
Here, the operator  $c^\dagger_{i\lambda\sigma}$ creates an electron on rung $i$ and leg $\lambda=1,2$ with spin $\sigma$, $n_{i\lambda\sigma}$ is the electron number operator and $\vec{S}_{i\sigma}=c^\dagger_{i\lambda}\frac{\vec{\sigma}}{2}c_{i\lambda}$. The Hilbert space excludes all states with double occupied sites.

The coupling to the external pairfields shown in Fig.~\ref{fig2} is given by
\begin{equation}
H_1 = \Delta_1\left(P^\dagger_1 + h.c. \right) + \Delta_2\left(e^{i\phi} P^\dagger_L + h.c.\right)\, ,
\label{h1}
\end{equation}
where
\begin{equation}
P^\dagger_1=\left( c^\dagger_{1,1\uparrow}c^\dagger_{1,2\downarrow}-c^\dagger_{1,1\downarrow}c^\dagger_{1,2\uparrow}\right) /{\sqrt{2}}
\end{equation}
creates a singlet pair on the first rung on the left end of the ladder and
\begin{equation}
P^\dagger_L=\left( c^\dagger_{L-1,1\uparrow}c^\dagger_{L,1\downarrow}-c^\dagger_{L-1,1\downarrow}c^\dagger_{L,1\uparrow}\right) / {\sqrt{2}}
\end{equation}
creates a singlet pair on the last leg section the right end of the ladder (see Fig.~\ref{fig2}b). In the following we will set $\Delta_1 = \Delta_2 = t = 1$ and use $\phi$ to control the Josephson pair current through the ladder. The geometry of the external pairfield connections is similar to that used by Van Harlingen in his study of cuprate superconductors, \cite{Van Harlingen} reproduced here in Fig.~\ref{fig2}a), and to the arrangements proposed by Sigrist and Rice \cite{sigrist and rice} 
In the `d-wave' like phase, we expect that there will be a $\pi$ phase shift associated with the sign difference between the rung and leg pairfields, as shown in Fig.~\ref{fig2}. As we will discuss, the sawtooth curves in Figs.\ref{fig2}c) and d) represent the perfect Andreev reflection boundary condition limit and the Josephson sine waves represent the response in the case in which the boundary conditions evolve to zero Andreev reflection.\cite{SHS} 

%The operator that measures the current flow along the ladder is

The Density Matrix Renormalization Group (DMRG) method \cite{dmrg} was used to calculate the ground state of the Hamiltonian $H+H_1$. The term (\ref{h1}) does not conserve the particle number, but it preserves the total spin, and charge even-odd parity (number of fermions modulo 2). In order to work at a fixed density, the chemical potential $\mu$ has to be carefully calculated for each system size. Working in the grand canonical ensemble, as well as using complex numbers due to the Josephson phases, clearly limits our ability to go to very large systems. We have typically studied ladders of up to $L=24$ rungs, keeping a total 600 DMRG states, with a truncation error or the order of $10^{-6}$, or smaller.

\begin{centering}
\begin{figure}
\epsfig {file=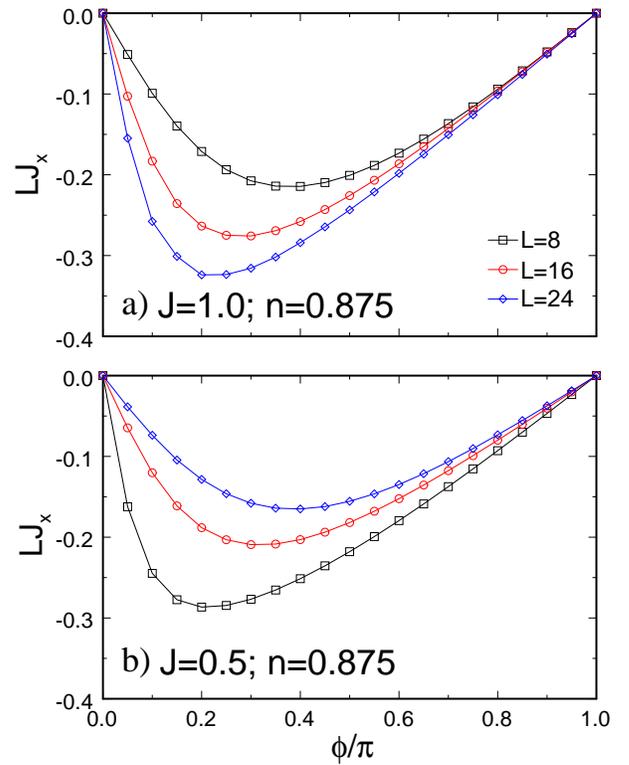,width=80mm}
\caption{Josephson current through $t-J$ ladders of different lengths $L$, at a fixed density $\langle n \rangle = 0.875$. The upper panel shows an increasing response which approaches a $\pi$-shifted d-wave sawtooth as $L$ increases for $J/t = 1.0$. In the lower panel, for $J/t = 0.5$, the response approaches a Josephson sine wave form and decreases as $L$ increases.
}
\label{fig3}
\end{figure}
\end{centering}

Away from the contacts, the current distribution is dominantly along the legs
\begin{equation}
J_x(i)=-it\sum\limits_{\sigma,\lambda} \langle c^\dagger_{i,\lambda\sigma} c^{\phantom{\dagger}}_{i+1,\lambda\sigma}-c^\dagger_{i+1,\lambda\sigma}
c^{\phantom{\dagger}}_{i,\lambda\sigma} \rangle\, , 
\end{equation}
and is independent of of the position of the rung $i$.  
Fig.~\ref{fig3}a) shows DMRG results for $L\langle J_x \rangle$ for $J/t=1.0$ and $\langle n \rangle = 0.875$ for ladders of various lengths $L$. As $L$ increases, the current versus the phase $\phi$ evolves into a $\pi$-phase shifted sawtooth curve. This is what one would expect for a 'd-wave' like superconductivity state. That is, just as discussed in Ref.[\onlinecite{Affleck}] for a one-dimensional electron liquid, as the length of the ladder increases the boundary condition flows to one with perfect Andreev reflection when the pairing correlations are dominant.
We previously found this in similar DMRG calculations for the one-dimensional Hubbard model \cite{SHS} with an attractive interaction. 
In the present 2-leg $t-J$ ladder, the difference is that the sawtooth is shifted by $\pi$ due to the d-wave nature of the pairing.

In Ref.[\onlinecite{Affleck}] bosonization and renormalization group methods were used to 
predict the scaling of the current with length, for a single leg Hubbard or 
$t-J$ model.  These methods can be readily extended to the two-leg case, based on 
the well-understood behavior of that model as studied, for example, in Ref.
[\onlinecite{friedel}]. The rung boundary term, at the left-hand side 
of the system adds a term:
\begin{equation}H_1=-C\Delta \cos (\sqrt{\pi}\varphi (0)),\end{equation}
to the bosonized Hamiltonian.  Here $C$ is a constant (which 
can be taken to be positive) and $\varphi (x)$ is the field arising
from bosonization associated with the total charge degrees of freedom 
which was labelled $\phi_{+\rho}$ in Ref.[\onlinecite{friedel}]. 
Taking into account the (normal) free end boundary condition (b.c.), 
$\theta (0)=$ constant (where $\theta$ is the field dual to
$\varphi$), we see that this interaction has a renormalization 
group scaling dimension of $1/(2K)$ where $K$ is the Luttinger 
parameter (referred to as $K_{+\rho}$ in Ref.[\onlinecite{friedel}]). 
This is relevant for $K>1/2$ and irrelevant for $K<1/2$. In the 
relevant case, it is plausible that the coupling constant $\Delta$ 
renormalizes to infinity, changing the b.c. to $\varphi (0)=0$, 
which corresponds to a perfect Andreev reflection b.c. 
We can test the consistency of this hypothesis by considering 
the lowest dimension boundary operator which could appear in the effective 
fixed point Hamiltonian. For the single leg case this is a 
normal reflection term $\psi^\dagger_L\psi_R$+ h.c.  However, 
for the 2-leg ladder in the 'd-wave-$4k_F$CDW' phase studied in Ref.[\onlinecite{friedel}] 
this operator has exponentially decaying correlation functions 
and is strongly irrelevant.  The most relevant boundary operator 
is $(\psi^\dagger_L\psi_R)^2+ h.c. \propto \cos (2\sqrt{\pi}\theta )$
 corresponding to normal pair reflection. 
This has a scaling dimension $2K$ and thus is relevant for $K<1/2$ 
and irrelevant for $K>1/2$.  This strongly suggests that 
the renormalization group flow is from normal to Andreev fixed points for $K>1/2$ 
and from Andreev to normal for $K<1/2$, just like in the single leg case.    

The Josephson current can now be calculated by scaling arguments 
using these results.  For $K<1/2$ the pairing interactions 
scale towards zero as $\Delta (L)\propto 
1/L^{1/(2K)-1}$, so we expect the current 
to scale to zero as $(1/L)\Delta (L)^2
\propto 1/L^{1/K-1}$ and to have an approximately 
sinusoidal form at large $L$. For $K>1/2$, as $L\to \infty$, we
expect to get a sawtooth current:
\begin{equation} J_xL=2vK(-1+\phi /\pi ),\ \  (0<\phi <2\pi )
\label{J}\end{equation}
where $v$ is the velocity of the gapless low energy charge excitations
 (referred to as $v_{+\rho}$ in Ref.[\onlinecite{friedel}]).
Here we have included the $\pi$-phase shift which arises from 
the fact that the leg pairing operator at the right end of the system
leads to a term like $H_1$ in the bosonized Hamiltonian but of 
opposite sign (and reduced magnitude). Eq. (\ref{J}) can be 
derived using $J=-2dE/d\phi$ and calculating $E(\phi )$ using 
the low energy effective Hamiltonian of Eq. (2.6) 
of Ref.[\onlinecite{friedel}] with boundary conditions 
$\varphi (0)=0$, $\varphi (L)=\sqrt{\pi}-\phi /\sqrt{\pi}$. 
The corrections to this sawtooth behavior scale to zero with $L$ 
and are controlled, at large $L$, by the leading irrelevant operator 
of dimension $2K$. For instance, the critical current should 
behave as:
\begin{equation}J_cL=2vK[1-V(L)^f],\end{equation}
where
\begin{equation}
V(L)\propto 1/L^{2K-1}
\end{equation}
is the renormalized normal pair reflection amplitude at scale $L$. 
The exponent $f$ was argued to be $f=2/3$ in
Ref.[\onlinecite{Affleck}] by consideration of the 
non-interacting case, but 
it might be expected to have a different 
value for the 2-leg ladder where the leading irrelevant 
operator corresponds to normal {\it pair} reflection rather than normal 
electron reflection.
We calculated $K$ and $v$ by studying the finite size scaling of the compressibility and the excitation gap for $S^z=0$, for system sizes up to $L=56$, and using 1500 DMRG states.\cite{friedel}
For $J/t=1$ and $\langle n \rangle = 0.875$, we find $v = 0.313$; $K = 0.766$; $2vK = 0.480$, which is in good agreement with the limiting behavior of $L\langle J_x \rangle$ seen in Fig.~\ref{fig3}a), that shows a slope of $0.484$.

From the phase diagram of Fig.~\ref{fig1}, one sees that the parameter $K$ decreases as $J/t$ is reduced.
 So, keeping the filling fixed at $\langle n \rangle = 0.875$, we studied the behavior of $L\langle J_x \rangle$ for smaller values of $J/t$. Fig.~\ref{fig3}b) shows $L \langle J_x \rangle$ versus $\phi$ for $J/t = 0.5$ for ladders of various lengths.
In this case, we find that the response maintains its $\pi$-phase shifted d-wave like character, but as $L$ increases it evolves to a Josephson like $\sin{(\phi+\pi)}$ form. In addition, the amplitude decreases. This is what one would expect if the boundary conditions associated with the pairfield coupling evolve to zero Andreev reflection as $L$ increases.

\begin{centering}
\begin{figure}
\epsfig {file=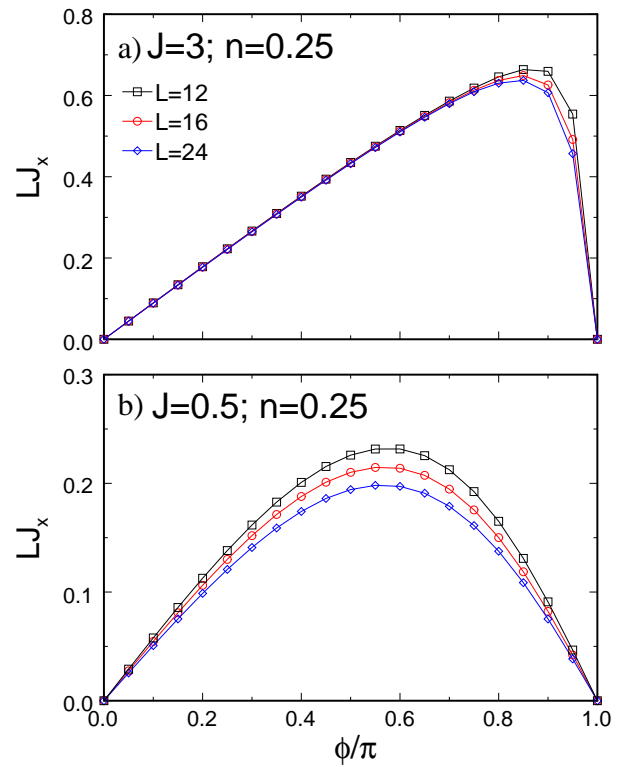,width=80mm}
\caption{Josephson current through $t-J$ ladders of different lengths $L$, at a fixed density $\langle n \rangle = 0.25$. The upper panel shows a sawtooth response for $J/t = 3.0$. In the lower panel, for $J/t = 0.5$, the response decreases with increasing system size, resembling a typical sine-like Josephson shape.
}
\label{fig4}
\end{figure}
\end{centering}

Based upon the phase diagram of Fig.~\ref{fig1}, we would have expected that one would have had to go to a lower value of $J/t$ for this to happen.  
In Ref.[\onlinecite{friedel}], the value of $K$ was estimated by two different methods, obtaining $K=0.604$ and $0.633$ for $J/t=0.5$ and
 $\langle n \rangle = 0.875$. This is in apparent contradiction with the renormalization group arguments given above and the behavior seen in Fig.~\ref{fig3}b). 
%From the phase diagram of Ref.[\onlinecite{phase diag}], calculated using the Drude weight, it appears that this point is very close to $K=0.5$. 
Thus, it may be that, for these parameters where $K$ is rather close 
to its critical value of $1/2$, larger system sizes must be studied to see the asymptotic behavior. 
Another possible explanation would be a non-trivial phase that exists for intermediate 
values of $\Delta$, for a range of $K$.

In contrast to the $\pi$-phase shifted 'd-wave' response, a similar set of calculations for the case $\langle n \rangle =0.25$, shown in Figs.\ref{fig4}a and b, clearly exhibit an 's-wave' response. For $J/t=3.0$, the $L\langle J_x \rangle$ curves exhibit the expected sawtooth form although the response appears essentially independent of system length $L$. For $J/t = 0.5$, when the CDW correlations are dominant, the flow is as expected to a Josephson $\sin{\phi}$ form associated with total normal reflection boundary conditions.
While this is an unphysical region of the $t-J$ ladder, it provides a further illustration of the way in which pairfield couplings can be used to explore the superconducting response of lattice models.

%In contrast to this `d-wave' response, a similar set of calculations for 
%$J/t=2.5$ and $\langle n \rangle = 0.25$ is shown in Fig.~\ref{fig2}b). Here 
%one is in an `s-wave' like region of the phase space and one sees that the 
%response goes towards the expected sawtooth form.
%
%Another characteristic behavior of $J_x$ versus $\phi$ is shown in 
%Fig.~\ref{fig2}c). Here for $J/t=0.2$ and $\langle n \rangle = 0.875$, 
%the system is in the `d-wave' region as seen by the $\pi$-phase shift. 
%However $\kappa_\rho < 0.5$ and $J_x$ exhibits the weak coupling 
%$-I\sin{\phi}$ Josephson behavior whose amplitude decreases as $L$ increases. 
%In this case, 
%the boundary conditions are renormalized towards the case of zero Andreev 
%reflection as the length $L$ of the ladder increases.

\begin{centering}
\begin{figure}
\epsfig {file=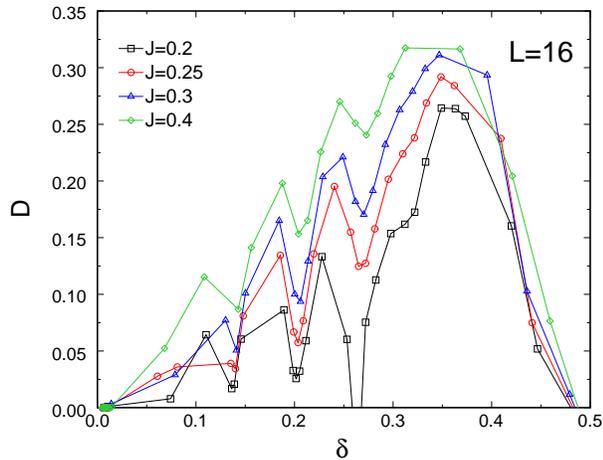,width=65mm,angle=-90}
\caption{Josephson current through a $t-J$ ladder with $L=16$, for different values of $J/t$.
$D$ is defined as the slope of the current $D=L|\langle J_x \rangle|/0.1\pi$, for
$\phi=0.9\pi$. 
We show results as a function of the hole concentration $\delta=1-\langle n \rangle$ in the d-wave region $\delta < 0.5$. All quantities are in units where the hopping $t=1$.
} 
\label{fig5} 
\end{figure} 
\end{centering}

Finally, for a finite ladder, it is interesting to study the behavior of $L\langle J_x \rangle/\phi$ versus the doping $\delta=1-\langle n \rangle$ for various values of $J/t$. As seen from the schematic phase diagram shown in Fig.~\ref{fig1}, at small values of $J/t$, one expects to cut through a CDW transition when $\delta=0.25$. We define the {\it effective} pair charge stiffness $D$ as the slope of the Josephson current near $\phi=\pi$. Results showing $D=L \langle J_x \rangle/(\phi-\pi)$ versus $\delta$ for several values of $J/t$ are plotted in Fig.~\ref{fig5}. As expected for small values of $J/t$, there is a sharp dip in $D$ as $\delta$ passes through $0.25$. This dip decreases as $J/t$ increases and one passes further away from the endpoint of the CDW line-phase.
There are also dips associated with shifts in the doping $\Delta \delta = 1/16$ corresponding to the addition of pairs. Smooth boundary conditions tend to remove these sharp features.\cite{sbc} Although one would need a detailed scaling study of $LJ_x/(\phi-\pi)$ to determine the superfluid weight, the dip at $\delta=0.25$ for $J/t=0.3$ clearly shows that the $t-J$ ladder exhibits a suppression in its superconductivity in the region near where static stripes appear.
Another aspect of this, which we have observed, is that when the boundary pairfield is relevant, the Friedel density oscillations generated by the open ends are reduced in amplitude.

In summary, by applying pairfields we have carried out the numerical equivalent of a phase sensitive experiment on $t-J$ ladders and studied the pair transport properties of a superconductor-Luther-Emery-superconductor (S-LE-S) junction. We have seen how the shape of the pair-current-phase characteristic can be used to determine the rung-leg parity of the superconducting pairs and its strength. One can also probe the effect of other phases on the superconducting response.

AEF and DJS would like to acknowledge insightful conversations with T. Giamarchi. SRW acknowledges the support of the NSF under grant DMR-0605444, and 
DJS would like to acknowledge the Center for Nanophase Material Science at Oak Ridge National Laboratory for support. IA acknowledges support from NSERC
 and CIAR.

\end{document}